\documentclass[12pt]{article}
\begin{document}
\newfont{\elevenmib}{cmmib10 scaled\magstep1}%
\renewcommand{\theequation}{\arabic{section}.\arabic{equation}}
\newcommand{\tabtopsp}[1]{\vbox{\vbox to#1{}\vbox to12pt{}}}

\newcommand{\preprint}{
            \begin{flushleft}
   \elevenmib Yukawa\, Institute\, Kyoto\\
            \end{flushleft}\vspace{-1.3cm}
            \begin{flushright}\normalsize  \sf
            YITP-99-73\\
   {\tt hep-th/0001074} \\ December 1999
            \end{flushright}}
\newcommand{\Title}[1]{{\baselineskip=26pt \begin{center}
            \Large   \bf #1 \\ \ \\ \end{center}}}
\newcommand{\Author}{\begin{center}\large \bf
           R. Caseiro${}^a$, J.-P. Fran\c{c}oise${}^b$ and R.
Sasaki${}^c$ \end{center}}
\newcommand{\Address}{\begin{center}
            $^a$ Universidade de Coimbra\\
     Departamento de Matem\'atica\\
     3000 Coimbra, Portugal \\
     $^b$ Universit\'e de Paris 6, UFR 920, tour 45-46, 4 place
     Jussieu, B.P. 172,\\
     Laboratoire ``G\'eom\'etrie Diff\'erentielle, Syst\`emes Dynamiques
     et Applications"\\
75252 Paris, France\\
     ${}^c$ Yukawa Institute for Theoretical Physics\\
     Kyoto University, Kyoto 606-8502, Japan
      \end{center}}
\newcommand{\Accepted}[1]{\begin{center}{\large \sf #1}\\
            \vspace{1mm}{\small \sf Accepted for Publication}
            \end{center}}
\baselineskip=20pt

\preprint
\thispagestyle{empty}
\bigskip
\bigskip

\Title{Algebraic Linearization
 of Dynamics  of \\ Calogero Type
 for any Coxeter Group}
\Author

\Address
\vspace{0.6cm}

\begin{abstract}
Calogero-Moser systems can be generalized for any root system (including the
non-crystallographic cases).
The algebraic linearization of the generalized
Calogero-Moser systems and of their quadratic (resp. quartic)
perturbations are discussed.
\end{abstract}
\bigskip
\bigskip

\renewcommand{\thesection}{\Roman{section}}
\section{Introduction}
\label{intro}
\setcounter{equation}{0}

The Calogero-Moser systems (\cite{Cal1}--\cite{Cal2}) were extended
to any semi-simple Lie algebras by Olshanetsky and Perelomov (\cite{OP})
at the classical level. It was later generalized by Bordner, Corrigan and
Sasaki (\cite{bcs}) to any root system (including the non-crystallographic
case).
The rational Calogero-Moser system displays a perturbation of special
interest whose solutions are
all periodic of the same period. Such a type of perturbation was considered
for any
root system by Bordner, Corrigan and Sasaki and they proved the existence of
a Lax
pair which generalizes the one introduced by Olshanetsky-Perelomov
(\cite{OP})
for the
$A_{m-1}$ case.
In this article, we follow the method introduced in
Caseiro-Fran\c{c}oise (\cite{cf}) to prove an explicit algebraic
linearization
of
several systems of Calogero-Moser or Ruijsenaars-Schneider type.
This techniques together with the general Lax matrix introduced by
Bordner-Corrigan-Sasaki allows to show the full periodicity of all
the orbits of the rational system perturbed by a confining
quadratic potential. A proof of the involution of the eigenvalues
of the Lax matrix is given by generalizing the original proof
by Fran\c{c}oise (\cite{jpf}) for the $A_{m-1}$ case.
We show next that the flow associated to each
eigenvalue
has all orbits periodic of the same period.

        In the third part of the article, we consider
perturbations of quartic type which are still integrable as shown
(in the $A_{m-1}$ case) by Fran\c{c}oise-Ragnisco (\cite{fr}) and we
show
the existence of a Lax pair for all Coxeter groups.
        In the fourth part, we extend the algebraic linearization
of the trigonometric or hyperbolic systems (Sutherland systems)
(\cite{Sut})
to any root system having the minimal representation.

        We postpone to further studies the semi-classical
analysis of the associated quantum systems.

\section{Superintegrability of the rational Calogero-Moser system for any
Coxeter group}
\label{supint}
\setcounter{equation}{0}
        The rational Calogero-Moser system for any Coxeter group was
first considered by Bordner, Corrigan and Sasaki (\cite{bcs}). They proved
the
classical integrability by constructing a universal Lax pair (see
also \cite{bst}--\cite{DHPH}). The
quantum case was discussed by Dunkl (\cite{Dunk}). In this article, we use
the techniques of algebraic linearization discussed in (\cite{cf}) for
the $A_{m-1}$ case. This yields the superintegrability of the
rational Calogero-Moser system for any Coxeter group.

Let us denote by $\Delta$ a root system of rank $r$. The
dynamical variables are (as before) the coordinates $q_{i},
i=1,...,r$ and their canonically conjugate momenta $p_{i},
i=1,...,r$.
        The Hamiltonian for the classical Calogero-Moser model is:

\begin{equation}
    {\cal H}={1\over2}p^2+
       {1\over2}\sum_{\rho\in\Delta_+}
          {g_{|\rho|}^{2}|\rho|^{2}\over{(\rho\cdot q)^2}},
          \label{(2.1)}
\end{equation}
in which the coupling constants $g_{|\rho|}$
 are defined on orbits of the corresponding
Coxeter group.
That is, for the simple Lie algebra cases
$g_{|\rho|}=g$ for all roots in simply-laced models
and  $g_{|\rho|}=g_L$
for long roots and $g_{|\rho|}=g_S$ for
short roots in non-simply laced models.
    Choose a representation ${\cal D}$ of dimension $D$ of the Coxeter group
(see
appendix), then define the $D{\times}D$ matrix:
\begin{equation}
     p\cdot\hat{H}:\quad (p\cdot\hat{H})_{{\alpha}{\beta}} =
    (p\cdot\alpha)\thinspace{\delta}_{{\alpha}{\beta}}, \label{(2.2)}
\end{equation}
where $\alpha$ and $\beta$ are vectors belonging to the
representation.

Introduce next the $D\times D$ matrices $X$, $L$ and $M$:
\begin{eqnarray}
    X&=&{\rm i}\sum_{\rho\in\Delta_{+}}g_{|\rho|}
       (\rho\cdot\hat{H})\thinspace{1\over{(\rho\cdot
       q)}}\thinspace\hat{s}_{\rho}, \label{(2.3)}\\
    L&=& p\cdot\hat{H}+X, \label{(2.4)}\\
M&=&{\rm i\over2}\sum_{\rho\in\Delta_{+}}g_{|\rho|}\thinspace
   {|\rho|^2\over{(\rho\cdot q)^{2}}}\thinspace\hat{s}_{\rho}, \label{(2.5)}
   \end{eqnarray}
and a diagonal matrix:
\begin{equation}
    Q=q\cdot\hat{H};\quad (Q)_{{\alpha}{\beta}} =
    (q\cdot\alpha)\thinspace{\delta}_{{\alpha}{\beta}}. \label{(2.6)}
\end{equation}

\noindent        The time evolution of the matrix $L$ along the flow of
the Hamiltonian displays the following equations:
{
\setcounter{enumi}{\value{equation}}
\addtocounter{enumi}{1}
\setcounter{equation}{0}
\renewcommand{\theequation}{\arabic{section}.\theenumi{\it\alph{equation}}}
\begin{eqnarray}
{\dot{L}}&=& [L,M], \label{(2.7a)}\\
\dot{Q}&=&[Q,M] + L. \label{(2.7b)}
\end{eqnarray}
\setcounter{equation}{\value{enumi}}
}
Introduce now the functions
\begin{eqnarray}
 F_{k}&=&{\rm Tr}(L^{k});\quad k=1,\ldots,D, \label{(2.8)}\\
G_{k}&=&{\rm Tr}(QL^{k});\quad k=0,\ldots,D-1, \label{(2.9)}
\end{eqnarray}
whose time-evolution displays:
\begin{eqnarray}
\dot{F}_{k}&=&0, \label{(2.10)}\\
\dot{G}_{k}&=& F_{k+1}. \label{(2.11)}
\end{eqnarray}
        This provides the algebraic linearization.

\newpage
{\bf Proposition \ref{supint}.1}

\bigskip
        The generalized Calogero-Moser system (\ref{(2.1)}) is
superintegrable for
any
Coxeter group.

{\bf Proof.}

        Introduce together with the $D$ first integrals $F_{k}$
the $D(D-1)/2$ extra first integrals $H_{k,k'}=F_{k}G_{k'}-F_{k'+1}G_{k-1}$.
Independent conserved quantities $F_{k}$ to be obtained from the
Lax equation (\ref{(2.7a)}) occur at such $k=1+{\bf exponent}$ of the
corresponding crystallographic root systems.
For the non-crystallographic root systems, they arise at
$(k=2, m)$
for the dihedral group $I_{2}(m)$,  $(k=2,6,10)$ for $H_{3}$ and
$(k=2, 12, 20, 30)$ for $H_{4}$. These are the degrees at which Coxeter
invariant polynomials exist (\cite{Hump}).

\section{Full periodicity of the confining potential case}
\label{period}
\setcounter{equation}{0}
    Integrable systems related to any Coxeter group first considered
 by Bordner, Corrigan and
Sasaki (\cite{bcs}) are obtained by adding to generalized rational
Calogero-Moser
systems
a confining potential. We recall here the notations and results of this
article concerned with the confining potential.

        The Hamiltonian is now:
\begin{equation}
    {\cal H}_{\omega}={1\over2}p^2+{1\over2}\omega^2q^2+
       {1\over2}\sum_{\alpha\in\Delta_+}
          {g_{|\alpha|}^{2} |\alpha|^{2}\over{(\alpha\cdot
          q)^2}}.
          \label{(3.1)}
\end{equation}
With the same matrices introduced above in the first paragraph,
the time evolution displays:
{
\setcounter{enumi}{\value{equation}}
\addtocounter{enumi}{1}
\setcounter{equation}{0}
\renewcommand{\theequation}{\arabic{section}.\theenumi{\it\alph{equation}}}
\begin{eqnarray}
{\dot{L}}&=& [L,M]-{\omega}^{2}Q, \label{(3.2a)}\\
\dot{Q}&=&[Q,M] + L. \label{(3.2b)}
\end{eqnarray}
\setcounter{equation}{\value{enumi}}
}
Introduce the matrices:
\begin{equation}
    L^{\pm}= L\thinspace {\pm}\thinspace {\rm i}\omega Q. \label{(3.3)}
\end{equation}
These matrices undergo the time evolution:
\begin{equation}
    \dot{L}^{\pm}={\pm}\thinspace {\rm i}\omega L^{\pm}+[L^{\pm},M].
\label{(3.4)}
\end{equation}

\noindent        It was then observed that the matrix
${\cal{L}}=L^{+}L^{-}$ defines Lax matrix for the system:
\begin{equation}
    \dot{{\cal L}}=[{\cal L},M]. \label{(3.5)}
\end{equation}
        Consider then the functions:
{
\setcounter{enumi}{\value{equation}}
\addtocounter{enumi}{1}
\setcounter{equation}{0}
\renewcommand{\theequation}{\arabic{section}.\theenumi{\it\alph{equation}}}
\begin{eqnarray}
F_{k}&=&{\rm Tr}(L^{+}{\cal L}^{k}), \label{(3.6a)}\\
G_{k}&=&{\rm Tr}(L^{-}{\cal L}^{k}). \label{(3.6b)}
\end{eqnarray}
\setcounter{equation}{\value{enumi}}
}
The time evolution yields:
{
\setcounter{enumi}{\value{equation}}
\addtocounter{enumi}{1}
\setcounter{equation}{0}
\renewcommand{\theequation}{\arabic{section}.\theenumi{\it\alph{equation}}}
\begin{eqnarray}
\dot{F}_{k}&=&{\rm i}\omega F_{k}, \label{(3.7a)}\\
\dot{G}_{k}&=&-{\rm i}\omega G_{k}. \label{(3.7b)}
\end{eqnarray}
\setcounter{equation}{\value{enumi}}
}

        Thus these functions provide the algebraic
linearization of the system. The existence of this algebraic
linearization relies on the fact that the Lax equation is
supplemented by an extra-equation which provides the full dynamics.
Another consequence of this extra-equation is the involution of
the first integrals displayed by the Lax pair.

        Indeed, the formal structure of the equations [(3.2), 
        (\ref{(3.5)})] is
the same for any Coxeter group. So the same line of arguments
developed in (\cite{jpf}) for the special case $A_{m-1}$ shows the
following:
(which is also of interest in the rational case)

\bigskip
{\bf Theorem \ref{period}.1}

\bigskip
        The Hamiltonian flows generated by the functions
$$H_{k}={\rm Tr}({\cal L}^{k}), \quad k=1,\ldots, D$$
Poisson commute.

{\bf Proof.}

        Consider the symplectic form:
\begin{equation}
    \Omega = {\rm Tr}(dQ{\wedge}dL)=C_{\cal D} \sum_{j=1}^{r}
    dq_{j}{\wedge}dp_{j}, \label{(3.8)}
\end{equation}
defined on the product of two copies of the representation. The
constant $C_{\cal D}$ depends actually on the representation. We can in
the following forget about this factorizing constant and simply
consider the Hamiltonian system defined by:
{%
\setcounter{enumi}{\value{equation}}
\addtocounter{enumi}{1}
\setcounter{equation}{0}
\renewcommand{\theequation}{\arabic{section}.\theenumi{\it\alph{equation}}}
\begin{equation}
\Omega = {\rm Tr}(dQ{\wedge}dL), \label{(3.9a)}
\end{equation}
and the Hamiltonian:
\begin{equation}
    {\cal H}_{\omega}=(1/2C_{\cal D})\thinspace{\rm Tr}({\cal L}).
\label{(3.9b)}
\end{equation}
\setcounter{equation}{\value{enumi}}
}%

        Let $\Lambda$ be an eigenvalue of the matrix $\cal L$ and let $T$
be the matrix of the projection onto the eigenspace corresponding
to this eigenvalue.
        Classical result of linear perturbation theory yields:
\begin{equation}
     d{\Lambda}= {\rm Tr}(d{\cal L}T). \label{(3.10)}
\end{equation}
        The Hamiltonian flow generated by the function $\Lambda$
and the symplectic form $\Omega$ displays:
\begin{equation}
    {\rm Tr}(\dot{Q}dL-\dot{L}dQ)= {\rm Tr}(d{\cal L}T). \label{(3.11)}
\end{equation}
This yields:
{
\setcounter{enumi}{\value{equation}}
\addtocounter{enumi}{1}
\setcounter{equation}{0}
\renewcommand{\theequation}{\arabic{section}.\theenumi{\it\alph{equation}}}
\begin{eqnarray}
\dot{Q}&=&[Q,M]+{\rm i}{\omega}[T,Q]+(LT+TL), \label{(3.12a)}\\
\dot{L}&=&[L,M]+{\rm i}{\omega}[T,L]-{\omega}^{2}(QT+TQ), \label{(3.12b)}
\end{eqnarray}
\setcounter{equation}{\value{enumi}}
}
and thus:
\begin{equation}
    \dot{\cal L}= [{\cal L},M]+2{\rm i}{\omega}[T,{\cal L}]
    =[{\cal L},M]. \label{(3.13)}
\end{equation}

        This shows that the eigenvalues of the Lax matrix
         ${\cal L}$ are constants of
motion for the Hamiltonian flow generated by any of its
eigenvalues.  In particular this proves that the Hamiltonian flows
generated by the eigenvalues of the Lax matrix ${\cal L}$ Poisson
commute.

\bigskip

{\bf Proposition \ref{period}.2}

\bigskip
        The Hamiltonian flows generated by any eigenvalues of the
Lax matrix ${\cal L}$ have all orbits periodic of the same period
${\pi}/{\omega}$.

{\bf Proof.}

        The time evolution along the Hamiltonian flow displays:
{%
\setcounter{enumi}{\value{equation}}
\addtocounter{enumi}{1}
\setcounter{equation}{0}
\renewcommand{\theequation}{\arabic{section}.\theenumi{\it\alph{equation}}}
\begin{eqnarray}
\dot{L}^{+}&=&[L^{+},M]+2{\rm i}{\omega}TL^{+}, \label{(3.14a)}\\
\dot{L}^{-}&=&[L^{-},M]-2{\rm i}{\omega}L^{-}T. \label{(3.14b)}
\end{eqnarray}
\setcounter{equation}{\value{enumi}}
}%
Introduce $U$ the (time dependent) matrix solution of the
Cauchy problem:
\begin{equation}
    \dot{U}=UM,{\quad} U(0)=1. \label{(3.15)}
\end{equation}
      The conjugated matrix $U{\cal L}U^{-1}$ is then a
constant of motion. Denote $V$ a time-independent matrix which
diagonalizes this matrix. Conjugate all the matrices $UL^{\pm}U^{-1}$
$UTU^{-1}$ by the matrix $V$ yields:
{%
\setcounter{enumi}{\value{equation}}
\addtocounter{enumi}{1}
\setcounter{equation}{0}
\renewcommand{\theequation}{\arabic{section}.\theenumi{\it\alph{equation}}}
\begin{eqnarray}
\dot{L}'^{+}&=& 2{\rm i}{\omega}{\tau}L'^{+}, \label{(3.16a)}\\
\dot{L}'^{-}&=&-2{\rm i}{\omega}L'^{-}{\tau}, \label{(3.16b)}
\end{eqnarray}
\setcounter{equation}{\value{enumi}}
}%
\noindent
where $L'^{\pm}=VUL^{\pm}U^{-1}V^{-1}$ and $\tau$ is the
constant diagonal matrix whose entries are equal to zero except
the diagonal term equals to $1$ in the position corresponding to
the eigenvalue.
        Equations (3.16) 
can be easily integrated and they yield
the periodicity of the eigenvalues of the matrix $Q=(1/2{\rm
i}{\omega})(L^{+}-L^{-})$ (the conservation of the Hamiltonian prevents
collisions).
 This clearly implies that the positions
$q$ are periodic in time of period ${\pi}/{\omega}$.

\section{The generalized rational Calogero-Moser system with an
external quartic potential}
\label{quartic}
\setcounter{equation}{0}

        The rational Calogero-Moser can be deformed into an
integrable system by adding a quartic potential (cf.\, (\cite{fr})).
We include now a proof of the existence of a Lax matrix for the
generalized rational Calogero-Moser system with an external
quartic potential.
        Define again the same matrices $L$, $Q$, $X$ and $M$. Let
$h(Q)=aQ+bQ^{2}$ be a matrix quadratic in $Q$; $(a,b)$ are just
two new independent parameters.
        The perturbed Hamiltonian is now (up to the normalization
constant $2C_{\cal D}$):
\begin{equation}
    {\cal H}_{h}\thinspace {\propto} \thinspace {\rm Tr}(L^{2}+h(Q)^{2}),
    \label{(4.1)}
\end{equation}

\bigskip
{\bf Theorem \ref{quartic}.1}

\bigskip
        The time evolution of the quartic type Hamiltonian system
(\ref{(4.1)}) can be cast into a Lax pair.

{\bf Proof.}

        Define the matrices:
\begin{equation}
    L^{\pm}=L\thinspace {\pm} \thinspace {\rm i} h(Q), \label{(4.2)}
\end{equation}
and
\begin{equation}
    {\cal L}=L^{+}L^{-}. \label{(4.3)}
\end{equation}
The time evolution equations (\ref{(3.4)}) of the matrices $L^{\pm}$
get modified as follows:
{%
\setcounter{enumi}{\value{equation}}
\addtocounter{enumi}{1}
\setcounter{equation}{0}
\renewcommand{\theequation}{\arabic{section}.\theenumi{\it\alph{equation}}}
\begin{eqnarray}
\dot{L}^{+}&=&[L^{+}, M-{\rm i}h'(Q)/2]+{\rm i}L^{+}h'(Q),
\label{(4.4a)}\\
\dot{L}^{-}&=&[L^{-}, M-{\rm i}h'(Q)/2]-{\rm i}h'(Q)L^{-}.
\label{(4.4b)}
\end{eqnarray}
\setcounter{equation}{\value{enumi}}
}%
This yields the Lax pair equation:
\begin{equation}
    \dot{\cal L}=[{\cal L},M-{\rm i}h'(Q)/2]. \label{(4.5)}
\end{equation}
        In the special limit $b=0$, $a=\omega$, the quartic system
reduces to the confining quadratic potential considered in the
paragraph \ref{period}.


\section{The trigonometric (hyperbolic) Calogero-Sutherland
system}
\label{trig}
\setcounter{equation}{0}
        The Hamiltonian of the trigonometric Calogero-Sutherland
model writes:
\begin{equation}
    {\cal H}={1\over2}p^2+
       {1\over2}\sum_{\alpha\in\Delta_+}
          {g_{|\alpha|}^{2}|\alpha|^{2}\over{\sin^{2}(\alpha\cdot q)}}.
          \label{(5.1)}
\end{equation}
        In order to get the hyperbolic case it suffices to change  $\sin$
into $\sinh$. In the following, we only demonstrate the algebraic
linearization of the trigonometric case. The hyperbolic case can
be deduced easily by the above replacement.
        We consider the matrices:
{
\setcounter{enumi}{\value{equation}}
\addtocounter{enumi}{1}
\setcounter{equation}{0}
\renewcommand{\theequation}{\arabic{section}.\theenumi{\it\alph{equation}}}
\begin{eqnarray}
L&=&p\cdot\hat{H}+X, \label{(5.2a)}\\
X&=&{\rm i}\sum_{\rho\in\Delta_{+}}g_{|\rho|}\thinspace
   (\rho\cdot\hat{H})\thinspace{1\over{\sin(\rho\cdot
   q)}}\thinspace\hat{s}_{\rho}, \label{(5.2b)}\\
M&=&-{\rm i\over2}\sum_{\rho\in\Delta_{+}}g_{|\rho|}\thinspace
   {|\rho|^2\cos(\rho\cdot q)\over{\sin^{2}(\rho\cdot q)}}
   \thinspace\hat{s}_{\rho},
\label{(5.2c)}
\end{eqnarray}
and diagonal matrices:
\begin{eqnarray}
Q&=&q\cdot\hat{H};\quad (Q)_{{\alpha}{\beta}} =
(q\cdot\alpha)\thinspace{\delta}_{{\alpha}{\beta}}, \label{(5.2d)}\\
R&=&{{\rm e}}^{2{\rm i}Q}. \label{(5.2e)}
\end{eqnarray}
\setcounter{equation}{\value{enumi}}
}

\newpage
{\bf Theorem \ref{trig}.1}

        In the case when the root system admits a minimal representation,
the time evolution along the flow of the Hamiltonian (\ref{(5.1)})
displays:
{
\setcounter{enumi}{\value{equation}}
\addtocounter{enumi}{1}
\setcounter{equation}{0}
\renewcommand{\theequation}{\arabic{section}.\theenumi\alph{equation}}
\begin{eqnarray}
\dot{L}&=&[L,M], \label{(5.3a)}\\
\dot{R}&=&[R,M]+{\rm i}(RL+LR). \label{(5.3b)}
\end{eqnarray}
\setcounter{equation}{\value{enumi}}
}%

        The algebraic linearization of the system (\ref{(5.1)}) follows
with the functions:
{
\setcounter{enumi}{\value{equation}}
\addtocounter{enumi}{1}
\setcounter{equation}{0}
\renewcommand{\theequation}{\arabic{section}.\theenumi\alph{equation}}
\begin{equation}
    a_{k}={\rm Tr}(L^{k}), \quad k=1,\ldots,D, \label{(5.4a)}
\end{equation}
and
\begin{equation}
    b_{k}={\rm Tr}(RL^{k}), \quad  k=1,\ldots,D, \label{(5.4b)}
\end{equation}
\setcounter{equation}{\value{enumi}}
}%
whose time evolution reads:
{
\setcounter{enumi}{\value{equation}}
\addtocounter{enumi}{1}
\setcounter{equation}{0}
\renewcommand{\theequation}{\arabic{section}.\theenumi\alph{equation}}
\begin{eqnarray}
\dot{a}_{k}&=&0, \label{(5.5a)}\\
\dot{b}_{k}&=&2{\rm i}b_{k+1}. \label{(5.5b)}
\end{eqnarray}
\setcounter{equation}{\value{enumi}}
}

{\bf Proof.}

        It can be easily checked that:
\begin{equation}
    \dot{R}= 2{\rm i}p\cdot\hat{H}R={\rm
    i}(p\cdot\hat{H}R+Rp\cdot\hat{H}), \label{(5.6)}
\end{equation}
\begin{equation}
    \dot{R}={\rm i}[(L-X)R+R(L-X)]={\rm i}(LR+RL)-{\rm i}(XR+RX).
    \label{(5.7)}
\end{equation}
We only need to show that:
\begin{equation}
    [R,M]=-{\rm i}(XR+RX). \label{(5.8)}
\end{equation}
Let us first evaluate the bracket $[R,M]$:
\begin{equation}
    [R,M]=-{\rm i\over2}\sum_{\rho\in\Delta_{+}}g_{|\rho|}\thinspace
       {|\rho|^2\cos(\rho\cdot q)\over{\sin^{2}(\rho\cdot q)}}
       \thinspace[{\rm e}^{2{\rm
    i}q\cdot\hat{H}
       }, \hat{s}_{\rho}]. \label{(5.9)}
\end{equation}
        The commutation relations:
\begin{equation}
    [\hat{H}_{j},\hat{s}_{\alpha}] = \alpha_{j}
       (\alpha^{\vee}\cdot\hat{H})\thinspace\hat{s}_{\alpha} \label{(5.10)}
\end{equation}
yield
\begin{equation}
    [R,M]=-{\rm i\over2}\sum_{\rho\in\Delta_{+}}g_{|\rho|}\thinspace
       {|\rho|^2\cos(\rho\cdot q)\over{\sin^{2}(\rho\cdot q)}}\thinspace{\rm
e}^{2{\rm
    i}q\cdot\hat{H}}
       (1-{\rm e}^{-2{\rm i}\rho\cdot{q}\rho^{\vee}\cdot\hat{H}
       })\thinspace\hat{s}_{\rho}. \label{(5.11)}
\end{equation}

\noindent
The minimal representation (cf. \cite{bcs}) is such that for all roots
$\rho$:
\begin{equation}
    \rho^{\vee}\cdot\hat{H}=0,{\pm}1. \label{(5.12)}
\end{equation}
For each fixed positive root $\rho$, three different cases have to be
considered:

\bigskip
i) In case $\rho^{\vee}\cdot\hat{H}=0$, then the right hand side
of the equation (\ref{(5.11)}) is zero. In this case, the contribution from
$XR+RX$ is zero as well because $X$ itself is zero.

ii) In case $\rho^{\vee}\cdot\hat{H}=1$, the contribution of
$\rho$ to the sum in (\ref{(5.11)}) reads:
\begin{equation}
    -{1\over2}g_{|\rho|}\thinspace
       {|\rho|^2\over{\sin(\rho\cdot q)}}\thinspace({\rm e}^{2{\rm
i}q\cdot\hat{H}}
       +{\rm e}^{2{\rm i}q\cdot\hat{H}-2{\rm
       i}\rho\cdot{q}\rho^{\vee}\cdot\hat{H}}
       )\thinspace\hat{s}_{\rho}. \label{(5.13)}
\end{equation}
       Since $\rho^{\vee}\cdot\hat{H}=1$, we have:
\begin{equation}
    (1/2)|\rho|^{2}={\rho}\cdot\hat{H}. \label{(5.14)}
\end{equation}
Then the above expression (5.13) reads:
\begin{equation}
    g_{|\rho|}{\rho}\cdot\hat{H}\thinspace
       {1\over{\sin(\rho\cdot q)}}\thinspace({\rm e}^{2{\rm
i}q\cdot\hat{H}}\hat{s}_{\rho}
       +\hat{s}_{\rho}{\rm e}^{2{\rm i}q\cdot\hat{H}}). \label{(5.15)}
\end{equation}
which is exactly the same as the contribution of ${\rho}$ to the
expression of $-{\rm i}(XR+RX)$.

iii) The third case $\rho^{\vee}\cdot\hat{H}=-1$ can be treated
analogously.

\section*{Acknowledgements}
\setcounter{equation}{0}
        The Authors warmly thank the organizers of NEEDS
conference where this research project started.
R. S. thanks the Universit\'e P.-M. Curie, Paris VI for hospitality.
        J.-P.F. is partially supported by a grant from the
French Ministry of Education and Research to the Laboratory
``G\'eom\'etrie diff\'erentielle, Syst\`emes dynamiques et
Applications", Universit\'e P.-M. Curie, Paris VI.
        R. S. is partially supported  by the Grant-in-aid from the
Ministry of Education, Science and Culture, Japan, priority area
(\#707) ``Supersymmetry and unified theory of elementary
particles".

\section*{Appendix. Root systems and finite reflection groups}

We now review some facts about root systems and their reflection
groups in order to introduce notation (cf. \cite{bcs}).  We
consider only
reflections in Euclidean space.
A root system $\Delta$ of rank $r$ is a set of
vectors in ${\bf R}^{r}$ which is invariant under reflections
in the hyperplane perpendicular to each
vector in $\Delta$:
$$
  \Delta\ni s_{\alpha}(\beta)=\beta-(\alpha^{\vee}\cdot\beta)\alpha,
   \quad \alpha^{\vee}={2\alpha\over{|\alpha|^2}},
\quad \alpha, \beta\in\Delta.
$$
        Once chosen a representation ${\cal D}$, the reflection is
represented by the operator $\hat{s}_{\alpha}$.

The set of positive roots $\Delta_{+}$ may be defined in terms of a
vector $V\in{\bf R}^{r}$, with
$V\cdot\alpha\neq 0,\,\forall\alpha\in\Delta$, as
those roots $\alpha\in\Delta$ such that $\alpha\cdot V>0$.

The set of reflections $\{s_{\alpha},\,\alpha\in\Delta\}$ generates a
group, known as a Coxeter group.

The root systems for finite reflection groups may be divided into two
types: crystallographic and non-crystallographic root systems.
Crystallographic root systems satisfy the additional condition
$$
   \alpha^{\vee}\cdot\beta\in{\bf Z},\quad \forall
 \alpha,\beta\in\Delta.
$$
These root systems are associated with simple Lie
algebras: $(A_{r},r\ge 1)$, $(B_{r},r\ge 2)$, $(C_{r},r\ge 2)$,
$(D_{r},r\ge 4)$, $E_{6}$, $E_{7}$, $E_{8}$, $F_{4}$, and
$G_{2}$ and  $(BC_{r},\,r\ge 2)$.  The Coxeter groups for these root
systems are called Weyl groups.  The remaining non-crystallographic root
systems are $H_{3}$, $H_{4}$, and the dihedral group of order $2m$,
$(I_{2}(m),m\ge 4)$.


\end{document}